\documentclass[apj]{emulateapj}

\newcommand{\samename}{\vrule height0.4pt depth0.0pt width1.0in \thinspace.}
\slugcomment{To appear in the Astrophysical Journal}

\shorttitle{AGB connection and UV excess in elliptical galaxies}
\shortauthors{Buzzoni \& Gonz\'alez-L\'opezlira}

\begin{document}

\title{AGB Connection and Ultraviolet Luminosity Excess in Elliptical Galaxies}

\author{Alberto Buzzoni\altaffilmark{1} and Rosa A. Gonz\'alez-L\'opezlira\altaffilmark{2}}
\email{alberto.buzzoni@oabo.inaf.it,\\ r.gonzalez@astrosmo.unam.mx}

\altaffiltext{1}{INAF - Osservatorio Astronomico di Bologna, Via Ranzani 1, 40127 Bologna (Italy)}
\altaffiltext{2}{Centro de Radioastronom\'\i a y Astrof\'\i sica, Universidad Nacional
Aut\'onoma de M\'exico, 58190 Morelia, Michoac\'an (Mexico)}

\begin{abstract}

Relying on infrared surface brightness fluctuactions to trace AGB
properties in a sample of elliptical galaxies in the Virgo and Fornax clusters,
we assess the puzzling origin of the ``UV-upturn'' phenomenon, recently 
traced down to the presence of a hot horizontal branch (HB) stellar component.
We find that the UV-upturn actually signals a profound change in the c-m 
diagram of stellar populations in elliptical galaxies, involving both the 
hot stellar component and red-giant evolution.  First, we encounter that the
strengthening of the UV rising branch is always seen to correspond to
a shortening in AGB deployment; this trend can be readily interpreted as an age
effect, perhaps mildly modulated by metal abundance.

A comparison between galaxy $\overline{K}$ magnitudes and population synthesis 
models confirms that, all the way, brightest stars in ellipticals are genuine AGB 
members, reaching the thermal-pulsing phase, and with the AGB tip exceeding the 
RGB tip by some 0.5-1.5~mag. The inferred core mass of these stars is found to be 
$\lesssim 0.57$~M$_\odot$ among giant ellipticals.
Coupled with the recognized severe deficiency of planetary nebulae in these galaxies,
this result strongly calls for an even more critical blocking effect due to a 
lengthy transition time needed by the post-AGB stellar core to become a hard UV emitter
and eventually ``fire up'' the nebula.

The combined study of galaxy $(1550 - V)_o$ color and integrated H$\beta$ index 
points, as an explanation for the UV-upturn phenomenon, to 
a composite HB with a bimodal temperature distribution, i.e.\ with both a red 
clump and an extremely blue component, in a relative proportion of roughly 
[N(RHB):N(BHB)]~$\sim$~[80:20]. As far as metallicity of the BHB stellar population is
concerned, we find that [Fe/H] values of either $\simeq -0.7$~dex or $\gtrsim +0.5$ may 
provide the optimum ranges to feed the needed low-mass stars (M$_* \ll 0.58$~M$_\odot$), 
that at some stage begin to join the standard red-clump stars.
\end{abstract}


\section{Introduction and theoretical framework}

The so-called ``UV-upturn'' phenomenon \citep{code79}, i.e., the rising
ultraviolet emission shortward of 2000 \AA, sometimes featuring in the 
spectral energy distribution (SED) of elliptical galaxies and the bulges of spirals,  
has been for long a puzzling problem for old galaxy environments dominated by 
stars of mass comparable to the Sun.

In fact, the implied existence of an important contribution of (long-lived) 
B stars, hotter than $\sim 30\,000$~K and providing up to
about 2\% of the galaxy bolometric luminosity \citep{rb86}, has been 
alternately identified with different evolutionary stages. 
Such stages include binaries \citep{brown06},
blue stragglers \citep{bailyn95}, 
blue horizontal-branch (HB) stars \citep{dorman95}, 
asymptotic giant branch (AGB) {\it manqu\'e} stars \citep{greggio},
and post-AGB nuclei of planetary nebulae (PNe) \citep{rb86}
\citep[see][, for an exhaustive review, and a more recent update by \citealp{yiyoon04}]{oconnell99}

Resolved color--magnitude (c-m) diagrams of  stellar populations in M32 
\citep{brown98,brown00} have definitely shown that even its relatively poor
UV emission almost entirely arises from a fraction of hot HB stars, further complemented
by a minority contribution from post-AGB PN nuclei. Still, facing the established 
interpretative scenario, one is left with at least three important issues 
that need to be assessed to understand the real nature of the UV-upturn phenomenon.

{\it (i)} 
The canonical evolutionary framework experienced in Galactic globular clusters
naturally predicts a blue HB morphology only for old, metal-poor stellar populations 
\citep{chiosi,rffp}. If this is the case for ellipticals too, then 
UV stars should represent the $Z \ll Z_\odot$ tail of a (supposedly) broad metallicity 
distribution seen to peak at much higher values, around solar abundance.
Clearly, a more composite picture might be envisaged 
once one admits non-standard models (i.e., including the effects of stellar rotation, 
helium mixing, differential mass loss, etc.) to account, in particular, for the well known 
``second parameter'' dilemma \citep{sweigart97,buonanno97,catelan01,recioblanco06}. 
However, this unconventional approach 
still suffers from a somewhat arbitrary fine-tuning of the key physical assumptions. 

{\it (ii)}
Hot HB stars might, nonetheless, 
also be naturally predicted among super metal-rich stellar models,
as far as metal abundance (and the linked helium content) exceeds some critical threshold. 
Presumably, in this case mass-loss allows stars to reach the HB phase
with a conveniently low external envelope, compared to the
helium core mass \citep{dorman,castellani92,yi,buzzoni95,dcruz96}. Such ``extreme HB'' stars (EHB)
have actually been observed, for example, in $\omega$~Cen \citep{dcruz00}, NGC~6388 and NGC~6441
\citep{rich97}, and in some old Galactic open clusters as well, like NGC~6791
\citep{kaluzny92,buson06}; they clearly remain the favorite candidates to explain the evolutionary 
framework of UV-enhanced elliptical galaxies \citep{moehler05}.

This hypothesis implies, however, a direct relationship between chemical abundance
(modulating the helium core mass at the HB onset) and mass-loss efficiency (to suitably ``peel off''
the stellar envelope along the RGB). As a consequence, one has to expect the UV-to-optical
color to be,
eventually, one of the most quickly evolving features in the SED of elliptical galaxies \citep{park97}.
In theory, the UV-upturn can fade by several magnitudes as the lookback time increases
by a few Gyr, although the effect is still detectable at intermediate redshift ($z \sim 0.3$)
\citep{brown03,ree06}. Unfortunately, the evolutionary details are extremely model-dependent, and a
strong UV excess could be triggered at ages as early as $\sim 6$ Gyr \citep{tantalo96} or as late 
as $\gtrsim$15 Gyr \citep{yi}.

{\it (iii)}
An established correlation seems to be in place  between
PN luminosity-specific rate and $(B-V)$ color for elliptical galaxies in 
the Virgo and Fornax clusters, and in the Leo group \citep{peimbert,hui}. The sense is that 
reddest metal-rich systems display, at the same time, a stronger UV-upturn
\citep{burstein88} {\it and} a poorer PN population per unit galaxy luminosity \citep{buzzoni06}.
If the PN event is the final fate for AGB stars at the end of their thermal pulsing phase 
\citep{ir83}, then the relative deficiency of nebulae might be evidence of an incomplete
(or fully inhibited) AGB evolution of low-mass stars under special environment conditions
of the parent galaxy.

As a central issue in this discussion, {\it it is clear therefore that a preeminent connection should
exist between UV excess and AGB distinctive properties of stellar populations in early-type 
galaxies.} 

On account of the Fuel Consumption Theorem \citep{rb86}, a 1~M$_\odot$ star
of solar metallicity enters its core He-burning phase with, 
at most, the equivalent of 0.43~M$_\odot$ of H to be spent as 
nuclear fuel.\footnote{Under the most extreme 
hypothesis of no mass-loss, a 1~M$_\odot$ star with solar abundance $(Y,Z) = (0.28,0.02)$ starts 
its HB evolution with a total He amount of roughly $0.62$~M$_\odot$, of which $\sim 0.47$~M$_\odot$
are confined in the core \citep{sg76} and $Y(1-0.47)\simeq 0.15$~M$_\odot$ reside in the envelope. 
Metals amount to roughly $Z(1-0.47) \simeq 0.01$M$_\odot$ and, accordingly, fresh H is 
0.37~M$_\odot$.
Taking into account the nuclear rates \citep[e.g.,][]{cox}, the H+He fuel provides
at most the equivalent of $0.37+0.62/10 \simeq 0.43$~M$_\odot$ of hydrogen.} 
This means that, under quite general conditions, post-RGB evolution alone could easily account, 
in principle, for up to 3/4 of the total bolometric luminosity of a galaxy stellar population 
\citep{buzzoni98}. Whether this energy is eventually reduced (if stars 
loose their fuel before they burn it), or whether it is finally released 
in the form of ultraviolet or infrared photons, crucially depends on mass-loss and its impact 
along the entire red-giant evolution. Hence, it is of special pertinence 
to constrain the relevant physical conditions that affect AGB
evolution in favor of an earlier transition of HB stars towards high temperature and enhanced 
ultraviolet emission.

In this paper we would like to draw the reader's attention to a possibly new and powerful 
approach to the problem, that can find straightforward applications even to distant galaxies.
As explained in Sec.~2, the method relies on surface-brightness fluctuation theory to safely tie 
infrared effective magnitudes (that {\it can} be determined for unresolved stellar populations), 
to stellar luminosity at the AGB tip (that {\it cannot} be directly observed in distant galaxies).
We will show, in Sec.~3, that these results tightly correlate with the ultraviolet properties 
of ``UV upturn'' elliptical galaxies, allowing a self-consistent physical picture and a quite accurate 
diagnostic of the post-RGB evolution of their underlying stellar populations, including HB 
morphology and AGB deployment. Our results will be finally summarized and discussed in Sec.~4.

\section{Infrared surface-brightness fluctuations as AGB probes}

\citet{tsch} and \citet{tonry91} first realized the potentially useful information about 
the composing stars hidden in the surface brightness fluctuations (SBFs) of galaxies with 
unresolved stellar populations. The problem has since received a more complete theoretical 
assessment by  \citet{buzzoni89,buzzoni93,buzzoni08,cervino00,cervino02}, and \citet{cervino06}. 

Briefly, the basic relationship upon which the theory relies is  
\begin{equation}
{{\sigma^2(L_{\rm gal})}\over {L_{\rm gal}}} = {{\sum \ell_*^2}\over {\sum \ell_*}} = \ell_{\rm eff}.
\label{eq:fluc}
\end{equation}
The l.h.\ side of the equation links an observable quantity (namely, the relative variance of the 
galaxy surface brightness) with the theoretical second-order statistical moment of the composing stars.
This quantity, in turn, traces the distribution of stellar luminosity ($\ell_*$) for the 
whole population; it is also a natural output of any population synthesis code,
and can easily be computed for different photometric bands and distinctive evolutionary
phases of simple stellar populations 
\citep[SSPs; see, e.g., the current theoretical databases of][]{buzzoni93,worthey93,blakeslee01,
cervino02,raimondo05,mouhcine05,cantiello03}.

The derived ``effective'' stellar luminosity, $\ell_{\rm eff}$, in eq.~(\ref{eq:fluc}) has
some important properties.
{\it (a)} It can be derived from a fully observational procedure, without any supplementary
theoretical assumptions.
{\it (b)} It is an intrinsic distinctive parameter of the stellar aggregate, 
and its empirical measurement does not require any 
individual star  to be (fully or partially) resolved.
{\it (c)} It identifies an ``effective'' magnitude (i.e. $\overline{M} = -2.5\log \ell_{\rm eff} +{\rm const}$)
in a completely similar way and with the same photometric zero points as ``standard'' magnitudes, 
and it scales accordingly with distance and foreground screen reddening.

\begin{figure}
\centerline{
\includegraphics[width=\hsize,clip=]{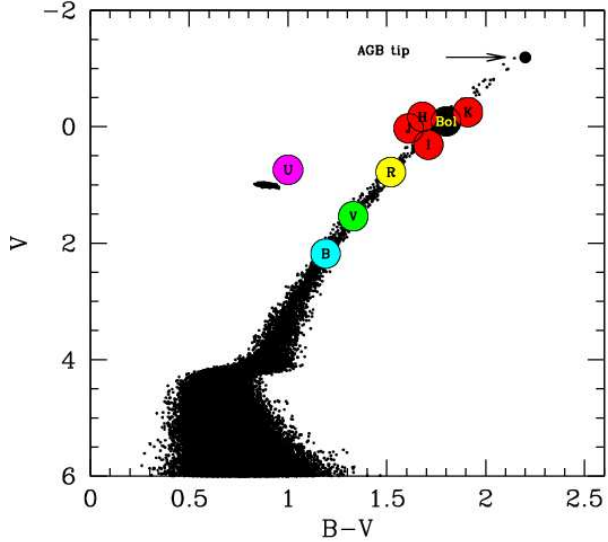}
}
\figcaption{
An illustrative  $V$ vs.\ $B-V$ synthetic c-m diagram for a 15 Gyr SSP of solar
metallicity and Salpeter IMF, with big dots identifying 
the different ``effective contributors'' (i.e., the appropriate magnitude 
$M \equiv \overline{M}$) at various photometric bands
(and to bolometric luminosity, as well). Although the different effective magnitudes cannot 
univocally be attributed to any $B-V$ color, one could let the prevailing stellar 
contributors to bolometric, infrared, and visual fluctuations arbitrarily coincide with the 
corresponding RGB location in the c-m diagram. This is not the case  
for the $U$ band, where the effective magnitude comes
from a more entangled mix of HB, red giants, and bright main sequence (MS) stars 
about the turn-off region of the diagram.
}
\label{cmd}
\end{figure}

\begin{figure*}
\centerline{
\includegraphics[width=0.492\hsize]{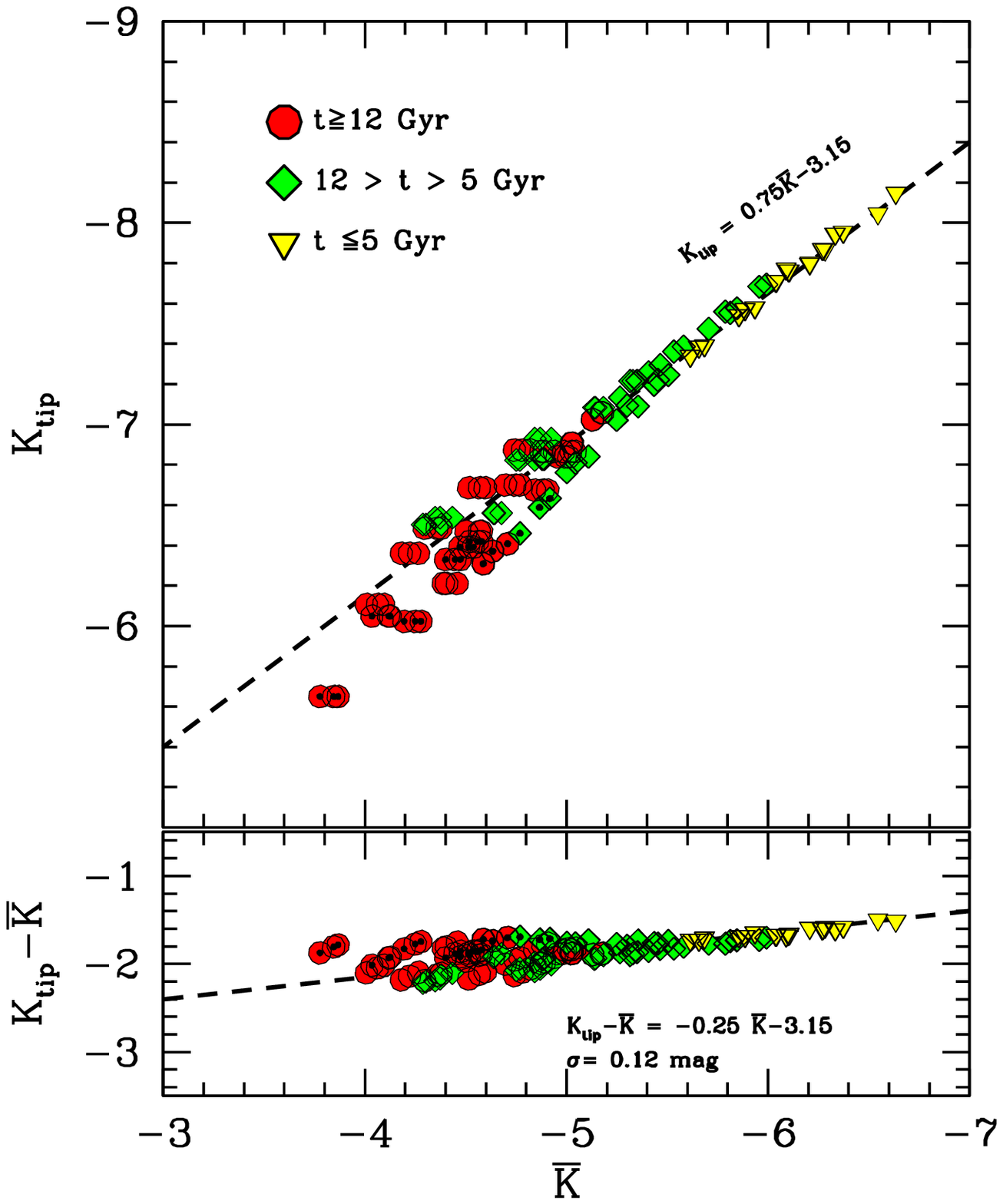}
\hspace{0.5cm}\includegraphics[height=0.585\hsize]{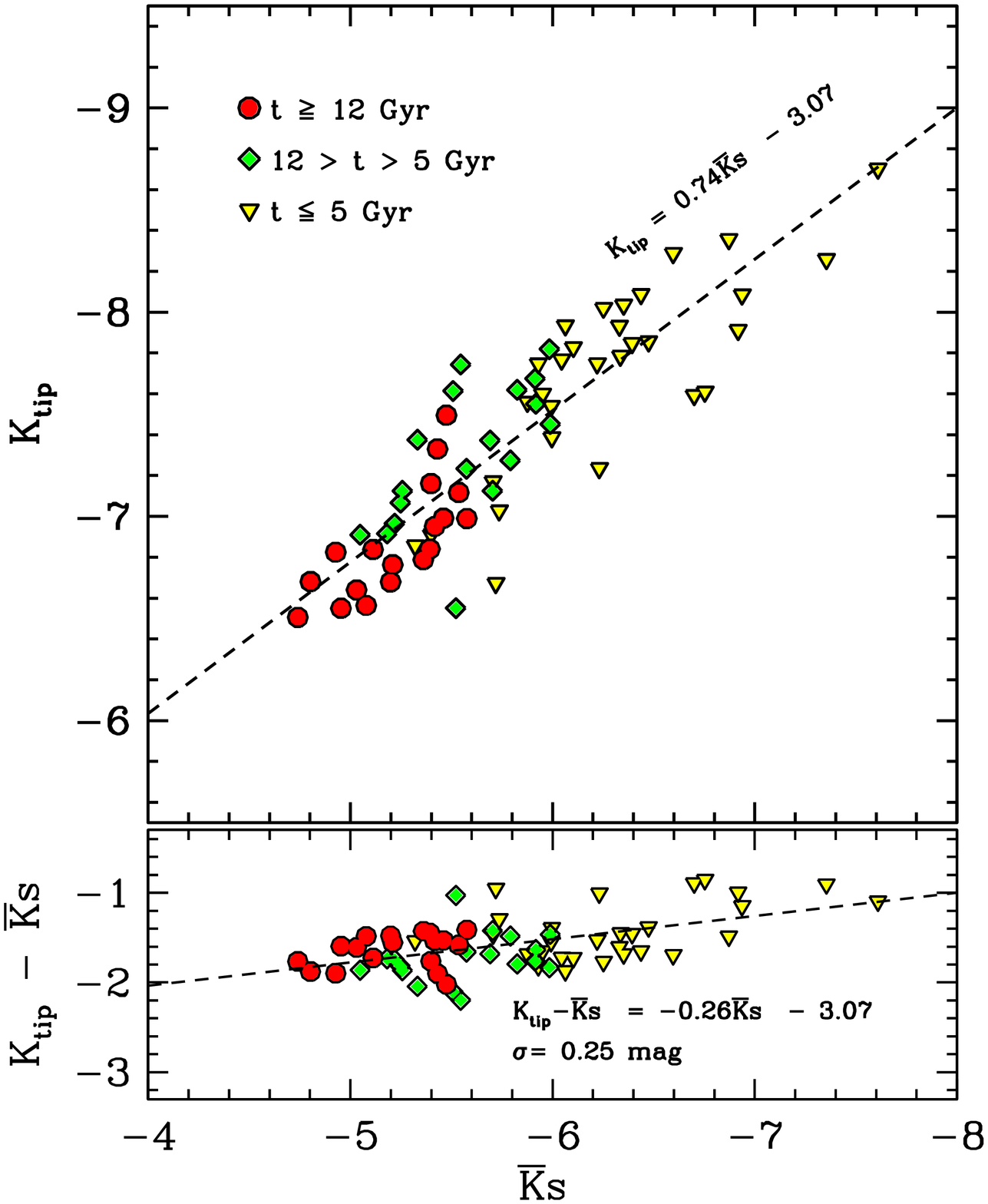}
}
\figcaption{
Theoretical relationship between near-IR effective magnitude and AGB+RGB tip luminosity, 
for a full collection of SSP models from \citet[][{\it, left panel}]{buzzoni89}, and 
Charlot \& Bruzual \citep[][{\it, right panel}]{gba07}.
Population ages span from 1 Gyr (upper right) to 18 Gyr (lower left); metallicity ranges
between $0.0001 \le Z \le 0.05$. The \citeauthor{buzzoni89} models explore a 
\citet{reimers} mass-loss parameter $\eta = 0.3$ and 0.5 for a Salpeter IMF,
while the Charlot \& Bruzual calculations adopt the mass-loss rates derived by \citet{mari07} 
and use a \citet{chab03} IMF. Dashed lines indicate fits to all points. Besides the 
overall agreement between the two synthesis codes, one has to notice a
larger scatter around the mean relation for the CB07 models, mainly due to a different response 
to SSP metallicity at younger ages.
}
\label{kk}
\end{figure*}

Being statistically representative of the stellar system as a whole, the effective magnitude cannot 
be physically associated to any specific star or stellar groups along the c-m diagram of a stellar
aggregate; nonetheless, it can be instructive to identify stars with $\overline{M}$ luminosity as the 
``prevailing'' group tracing the whole population in the different photometric bands. 
This is shown in Fig.\ 1 for the illustrative case of a 15 Gyr SSP with solar metallicity and Salpeter 
IMF.  Because of the quadratic $\ell_*$ dependence of the summation in the 
numerator of eq.~(\ref{eq:fluc}), 
the effective magnitude at any optical/infrared band is highly sensitive to the giant stars
and only marginally responds to a shift in the SSP ``mass leverage'' caused by a change in the IMF 
slope \citep[][]{buzzoni93}. 
Likewise, as probe of the brightest stars in a population at a given wavelength, 
the effective magnitude is relatively insensitive to any underlying older component 
in the case of composite stellar systems.\footnote{This fact has been observationally confirmed through
the comparison between SBF measurements in early-type galaxies and Magellanic
star clusters \citep[see][]{gonz04,gonz05a}.} 
 
It is evident from Fig.~1 that near-infrared (near-IR) magnitudes closely trace the bright-end
tail of the SSP luminosity function. In particular, one can notice that
$\overline{K}$ is potentially the best tracer of the SSP tip stellar luminosity ($K_{\rm
tip}$), as {\it both quantities are expected to depend in quite the same way on the overall
distinctive parameters of the stellar population, including age, metallicity, IMF, 
and mass-loss.}\footnote{In spite of probing 
a relative minority of bright giant stars, the infrared effective magnitude is a fully robust 
characteristic parameter of a stellar population, in the case of galactic mass scales.
From a statistical point of view, in fact, taking the reference SSP of Fig.~1 as
representative of the elliptical galaxy stellar population, one has to expect about $9{ \times}10^7$ 
``effective contributors'' (in the statistical definition of \citealp{buzzoni93}) to the $K$-band 
SBF for a galaxy of $10^{11}$~M$_\odot$. This number further increases to about $10^8$ for 
$H$-band contributors, and to $\sim 3{ \times}10^8$ for the $J$ band SBF.}
This is shown in Fig.~2, where we compare the $\overline{K}$ vs. $K_{\rm tip}$
relationship for a full collection of SSP models from the \citet{buzzoni89} and the updated
Charlot \& Bruzual \citep[CB07][]{gba07} synthesis codes.\footnote{The $K_s$ 
band is a variant of the standard $K$ filter, sometimes preferred for its reduced background noise 
\citep[see][ for specific details]{pers98}. Both filters have similar effective wavelength and 
photometric zero points, so that $K = K_s$ within a typical $\pm 0.03$~mag 
uncertainty, as we directly verified in our tests.}$^,$\footnote{For the CB07 models, the 
fluctuation luminosity has been computed via eq.~(\ref{eq:fluc}) from the original isochrone set, 
while $K_{\rm tip}$ is just read as the absolute magnitude of the 
brightest stellar type in the AGB+RGB phases.}

The correlation appears quite robust, in spite
of the extremely wide range explored for the population parameters, and the different input physics 
adopted by the two sets of theoretical models.

Formal fits to the whole sets of models yield: 

\begin{equation}
\left\{
\begin{array}{ll}
K_{\rm tip}  = 0.75\ \overline{K} - 3.15 \quad &\sigma = \pm 0.12~{\rm mag}\\
\\ 
 & \quad\qquad\qquad {\rm (Buzzoni~1989),}  \\ 
 & \\ 
K_{\rm tip} = 0.74\ \overline{K}_s - 3.07 \quad &\sigma = \pm 0.25~{\rm mag}\\  
\\ 
 & \quad\qquad\qquad {\rm (Bruzual~2007).} 
\end{array}
\right.
\label{eq:ddb}
\end{equation}

\subsection{AGB completion in young stellar populations: the case of Magellanic star clusters}

A comparison with real stellar populations extending across the widest range of evolutionary parameters
would clearly be the ideal test for the envisaged theoretical picture.
The Magellanic Clouds (MCs) stand out as striking candidates, in this regard, with the age of their 
star clusters spanning over four orders of magnitude, from a few Myr up to $\sim 10^{10}$~yr. 
In addition, their relatively close distance allows {\it both} fluctuation 
luminosity {\it and} AGB/RGB luminosity tip to be directly measured on resolved c-m diagrams. 

Near-IR SBFs for a sample of 191 MC star clusters have
been obtained by \citet{gonz04,gonz05} using $K_s$-band 
data from the Two Micron All Sky Survey \citep[2MASS,][]{skru97}. 
To reduce stochastic effects due to small-number
stellar statistics along fast evolutionary phases, in their study these authors 
assembled seven ``super-clusters'', by homogeneously co-adding several objects for each age 
class according to the \citet[][, SWB]{sear80} classification scheme.

The procedure to obtain the SBFs for these data has been extensively described before
\citep{gonz04,mouhcine05}; briefly, direct summation of bright stars provided the numerator 
of the r.h.\ fraction of eq.~(\ref{eq:fluc}), while total luminosity (at the denominator 
of the equation) was safely estimated from integrated photometry of each supercluster mosaic,
to include the contribution of faint, unresolved stars.
Following \citet{lee93}, for the present work we have determined the tip luminosity of the
AGB+RGB by convolving the luminosity function of each supercluster
with a Sobel edge-detection filter (the zero-sum kernel [-2,0,+2]).
The tip luminosity was identified with the peak of the filter response
function, after checking that there is indeed an important count
discontinuity at that location in the luminosity function; the error
in the measurement equals the width of the histogram bin. We list in
Table \ref{tab_mc} the $K_s$-band absolute fluctuation magnitude,
$K_s$ absolute tip magnitude, age, and metallicity of the MC
superclusters.

\begin{deluxetable}{lcccc}
\tablecaption{Parameters of MC stellar ``superclusters"}
\tablewidth{\hsize}
\tablehead{
\colhead{SWB}&\colhead{$\overline M_{K_s}$}
&\colhead{$M_{K_{\rm tip}}$}&\colhead{$\log$ age (yr)}&\colhead{$\log Z$}\\
\colhead{type}&\colhead{[mag]}&\colhead{[mag]}&\colhead{}&\colhead{}\\
\vspace*{-0.6cm}\\
}
\startdata
pre & --7.70$\pm$0.40&--8.7 $\pm$ 0.2 & 6.4 $\pm$ 0.3 & --0.3 $\pm$ 0.2\\
I   & --8.85$\pm$0.12&--9.8 $\pm$ 0.2 & 7.0 $\pm$ 0.3 & --0.3 $\pm$ 0.2\\
II  & --7.84$\pm$0.28&--8.9 $\pm$ 0.2 & 7.5 $\pm$ 0.3 & --0.3 $\pm$ 0.2\\
III & --7.45$\pm$0.24&--8.3 $\pm$ 0.2 & 8.0 $\pm$ 0.3 & --0.1 $\pm$ 0.2\\
IV  & --7.51$\pm$0.18&--8.4 $\pm$ 0.2 & 8.5 $\pm$ 0.3 & --0.8 $\pm$ 0.2\\
V   & --6.69$\pm$0.20&--8.0 $\pm$ 0.2 & 9.0 $\pm$ 0.3 & --0.6 $\pm$ 0.2\\
VI  & --6.21$\pm$0.24&--7.6 $\pm$ 0.2 & 9.5 $\pm$ 0.3 & --1.0 $\pm$ 0.2\\
VII & --4.92$\pm$0.38&--7.2 $\pm$ 0.2 & 9.9 $\pm$ 0.3 & --1.4 $\pm$ 0.2\\
\enddata
\tablecomments{
Col.\ (2): $\overline M_{K_s}$ from \citet{gonz04,gonz05}.
Col.\ (4): Ages from \citet{frog90}, corrected to the Large Magellanic Cloud 
distance modulus, ${\rm (m-M)_{\circ}=18.5}$, as in \citet{mouhcine05}; the age 
of the pre-SWB supercluster is also from \citet{gonz04}. 
Col.\ (5): Metallicities from \citet{cohe82}, for cluster types pre-SWB, 
I, and II, and from \citet{frog90} for later types.
}
\label{tab_mc}
\end{deluxetable}

\begin{figure}
\centerline{
\includegraphics[width=0.8\hsize]{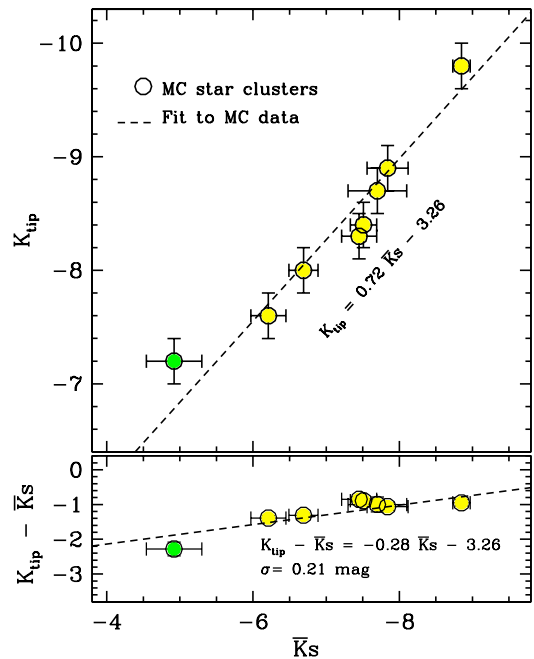}
}
\figcaption{Absolute $K_{\rm tip}$ vs.\ $\overline{K_s}$ magnitude relationship  
for Magellanic Cloud star clusters. Big dots locate the average position for 
each ``supercluster'' of Table~1, i.e.\ a co-addition of several individual systems 
belonging to homogenous \citet{sear80} age classes.
The dashed line is the fit to the data, as displayed on the plot. Color code for 
markers is like in Fig.~2.
}
\label{mcfig}
\end{figure}

Again, a tight relationship between $K_{\rm tip}$ and effective $\overline{K}$
magnitudes is in place, as shown in Fig.~3, along the entire age range
(and the corresponding metallicity drift, see column 5 in Table~\ref{tab_mc}). A fit to 
the MC data (dashed line in the figure) provides:

\begin{equation}
 K_{\rm tip} = 0.72\ \overline{K}_s - 3.26,\\
\label{eq:mc}
\end{equation}

\noindent
with a data scatter $\sigma = \pm 0.21$~mag across the fitting line. 
Note that this perfectly compares, within the statistical uncertainty of the observations,
with both eqs.~(\ref{eq:ddb}); this is true even beyond the nominal age range of the 
theoretical relations since, for instance, the pre-SWB and
the SWB type I MC superclusters are too young to even sport any standard AGB or RGB phase!

\subsection{Mass-loss and post-RGB evolution in evolved stellar populations}

The link between $\overline{K}$ and $K_{\rm tip}$ can be regarded as a {\it much 
more deeply intrinsic property of SSPs}, not exclusively related to age (or metallicity) 
evolution. Of course, our method cannot remove, by itself, any 
age/metallicity degeneracy, that makes integrated photometric properties of young 
metal-rich stellar systems closely resemble those of older metal-poor ones 
\citep[e.g.][]{rb86,worthey94}. However, as far as old stellar populations are concerned,
the unbiased and model-independent estimate of the tip stellar luminosity, even in unresolved 
stellar populations, provides an additional and powerful tool to quantitatively 
size up important effects, like the mass-loss impact on red-giant evolution.

\begin{figure}
\centerline{
\includegraphics[width=0.85\hsize]{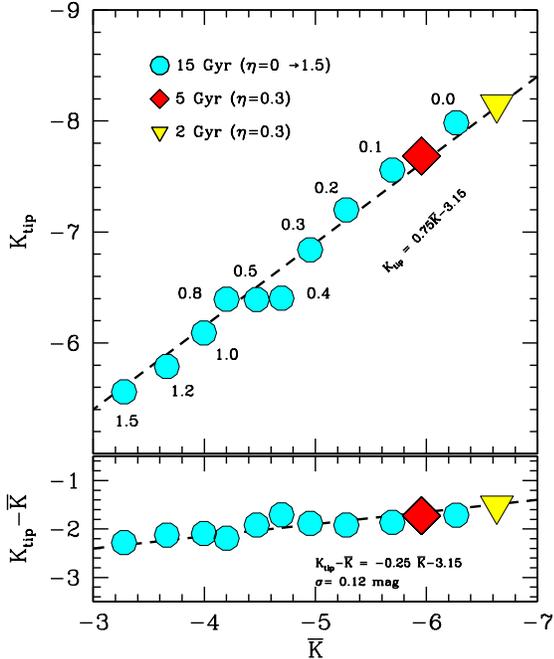}
}
\figcaption{
Theoretical relationship between $K_{\rm tip}$ and $\overline K$ with varying mass loss
efficiency along red-giant evolution. Big dots trace 
the change in a 15 Gyr old SSP of solar metallicity \citep{buzzoni89}, with 
the increase of Reimers mass-loss parameter $\eta$ from 0 to 1.5, as labelled on the
plot. Note that a reduced mass loss ($\eta \rightarrow$ 0) leads old SSPs to fully deploy
the AGB and closely resemble much younger 2-5 Gyr old SSPs (triangles and 
diamond markers on the plot) with $\eta = 0.3$ and otherwise same distinctive parameters.
}
\label{mloss}
\end{figure}

Relying on the \citet{reimers} standard $\eta$ parameterization, Fig.~4 gives an instructive 
example of how mass-loss can modulate
both $K_{\rm tip}$ and $\overline{K}$. In the figure we explore the behavior of an old 
(15 Gyr) SSP reference model (big dots in the plot), along a full variety of mass-loss scenarios,
spanning from a virtually vanishing stellar wind (i.e.\ $\eta = 0$), up to a violent mass-loss 
rate ($\eta \to 1.5$) capable of fully wiping out the external envelope of stars while on 
the RGB phase. Similarly to Fig.~3, we still report the same clean relationship between 
$K_{\rm tip}$ and $\overline{K}$, with old AGB-enhanced models ($\eta \to 0$) closely 
resembling, all the way,  young (2-5~Gyr) standard ($\eta \simeq 0.3$) populations in the 
upper right corner of the panel. 

Again, according to the trend of the 15 Gyr model sequence vs.\ $\eta$ we can propose the 
following parametric equations:

\begin{equation}
\left\{
\begin{array}{l}
K_{\rm tip}  \simeq 1.7\, \eta -7.7\\
\\
\overline{K} \simeq 2.1\, \eta -5.9. 
\end{array}
\label{eq:eta}
\right.
\end{equation}

\begin{deluxetable*}{rcccccccc}
\tabletypesize{\scriptsize}
\tablecaption{Summary of relevant data for the elliptical-galaxy sample}
\tablewidth{0pt}
\tablehead{
\colhead{NGC} & \colhead{$\log \sigma_{\rm v}$} & \colhead{Mg$_2 \pm \sigma$} & \colhead{$(B-V)_o$}
& \colhead{$\log \alpha$} & \colhead{$(1550-V)_o$} & \colhead{$\overline{F160W} \pm \sigma$} & 
\colhead{$\overline{K}_s \pm \sigma$} & \colhead{$H\beta \pm \sigma$} \\
\colhead{ } & \colhead{[km\,s$^{-1}$]} & \colhead{[mag]} & \colhead{[mag]}
& \colhead{ N$_{\rm PN}$\,L$_{\rm gal}^{-1}$} & \colhead{[mag]} & \colhead{[mag]} & \colhead{[mag]} & \colhead{[\AA]}}
\startdata
\vspace{0.8mm}
221   &  1.90 $\pm$ 0.03  &  0.198 $\pm$  0.007  & 0.88  & --6.77$^{0.18}_{0.31}$ & 4.50 $\pm$ 0.17  & --5.28 $\pm$ 0.10  &   --5.95 $\pm$ 0.14  & 2.30 $\pm$ 0.05 \\	      
\vspace{0.8mm}
224   &  2.27 $\pm$ 0.03  &  0.324 $\pm$  0.007  & 0.95  & --6.94$^{0.15}_{0.22}$ & 3.51 $\pm$ 0.17  & --4.46 $\pm$ 0.09  &   --5.69 $\pm$ 0.14  & 1.66 $\pm$ 0.07 \\
\vspace{0.8mm}
1316  &  2.38 $\pm$ 0.08  &  0.265 $\pm$  0.023  & 0.87  & --7.50$^{0.07}_{0.07}$ & 5.0  $\pm$ 0.2   & --5.39 $\pm$ 0.19  &       ...            & 2.20 $\pm$ 0.07 \\
\vspace{0.8mm}
1379  &  2.11 $\pm$ 0.13  &  0.269 $\pm$  0.006  & 0.88  &    ...                 & 3.01 $\pm$ 0.10  & --5.11 $\pm$ 0.19  &   --5.75 $\pm$ 0.12  & 1.70 $\pm$ 0.09 \\	   
\vspace{0.8mm}
1387  &    ...            &    ...               & 0.98  &    ...                 & 2.16 $\pm$ 0.05  & --5.4  $\pm$ 0.8   &  --5.77 $\pm$ 0.10   &   ...	    \\
\vspace{0.8mm}
1389  &  2.12 $\pm$ 0.06  &  0.236 $\pm$  0.002  & 0.90  &    ...                 & 3.38 $\pm$ 0.09  & --5.16 $\pm$ 0.20  &   --6.35 $\pm$ 0.13  &   ...	    \\
\vspace{0.8mm}
1399  &  2.52 $\pm$ 0.03  &  0.357 $\pm$  0.007  & 0.95  & --7.30$^{0.07}_{0.07}$ & 2.05 $\pm$ 0.17  & --4.55 $\pm$ 0.16  &   --5.28 $\pm$ 0.15  & 1.41 $\pm$ 0.08 \\
\vspace{0.8mm}
1404  &  2.39 $\pm$ 0.03  &  0.344 $\pm$  0.007  & 0.95  &    ...                 & 3.30 $\pm$ 0.17  & --4.76 $\pm$ 0.21  &   --5.53 $\pm$ 0.10  & 1.58 $\pm$ 0.08 \\
\vspace{0.8mm}
3379  &  2.33 $\pm$ 0.03  &  0.329 $\pm$  0.007  & 0.94  & --6.77$^{0.03}_{0.03}$ & 3.86 $\pm$ 0.17  & --4.70 $\pm$ 0.14  &   --5.43 $\pm$ 0.17  & 1.46 $\pm$ 0.16 \\
\vspace{0.8mm}
3384  &  2.20 $\pm$ 0.11  &  0.296 $\pm$  0.014  & 0.91  & --6.42$^{0.10}_{0.10}$ & 3.9 $\pm$ 0.2    & --4.82 $\pm$ 0.22  &     ...              & 2.05 $\pm$ 0.11 \\
\vspace{0.8mm}
4278  &  2.45 $\pm$ 0.03   &  0.293 $\pm$  0.007  & 0.90  &   ...                  & 2.88 $\pm$ 0.17  & --4.49 $\pm$ 0.22 &     ...              & 1.37 $\pm$ 0.03 \\
\vspace{0.8mm}
4374  &  2.48 $\pm$ 0.03  &  0.323 $\pm$  0.007  & 0.94  & --6.77$^{0.10}_{0.10}$ & 3.55 $\pm$ 0.17  &    ...             &   --5.71 $\pm$ 0.25  & 1.70 $\pm$ 0.04 \\
\vspace{0.8mm}
4406  &  2.42 $\pm$ 0.03  &  0.330 $\pm$  0.007  & 0.90  & --6.89$^{0.10}_{0.10}$ & 3.72 $\pm$ 0.17  &    ...             &   --5.74 $\pm$ 0.12  &  1.61$\pm$ 0.16 \\	  
\vspace{0.8mm}
4472  &  2.49 $\pm$ 0.03  &  0.331 $\pm$  0.007  & 0.95  & --7.16$^{0.10}_{0.10}$ & 3.42 $\pm$ 0.17  & --4.64 $\pm$ 0.11  &   --5.64 $\pm$ 0.13  & 1.52 $\pm$ 0.14 \\
\vspace{0.8mm}
4486  & 2.60 $\pm$ 0.03   &  0.303 $\pm$ 0.007   & 0.93  & --7.10$^{0.1}_{0.10}$  & 2.04 $\pm$ 0.17  &    ...             &      ...             & 1.38 $\pm$ 0.04 \\ 
\vspace{0.8mm} 
4552  & 2.44 $\pm$ 0.03   & 0.346  $\pm$  0.007  & 0.94  &    ...		  & 2.35 $\pm$ 0.17  &   ...              &  --5.86 $\pm$ 0.13   &  1.65$\pm$ 0.04 \\
\vspace{0.8mm}
4621  & 2.41 $\pm$ 0.03   & 0.355  $\pm$  0.007  & 0.92	 &	...		  & 3.19 $\pm$ 0.17  &   ...        	  &  --5.59 $\pm$ 0.21   & 1.43 $\pm$ 0.11 \\
\vspace{0.8mm}
4649 & 2.56 $\pm$ 0.03    &  0.360 $\pm$ 0.007   & 0.95  & --7.22$^{0.1}_{0.10}$  & 2.24 $\pm$ 0.17  &   ...              &      ...             & 1.26 $\pm$ 0.04 \\ 
\enddata
\tablecomments{ \scriptsize
Cols.~(2) and (3). Log $\sigma$ and Mg$_2$ index from \citet{burstein88}, except for NGC~1316, 
1379, 1389, 3384, that are averages from the Hyperleda database \citep{paturel}. 
Col.\ (4). ($B - V$)$_o$ from the RC3 Catalogue \citep{rc3}. 
Col.\ (5). Log~$\alpha$ from \citet{buzzoni06}. 
Col.\ (6) ($1550 - V$) color from \citet{burstein88}, except for NGC~1379, 1387, 1389, and 1404; 
these were scanned from \citet{lee05} with the Dexter package \citep{dexter}, and transformed according to the \citet{colina96}
calibration scale. Col.\ (7). $\overline{M}_{F160W}$ from \citet{jens03}.
Col.\ (8). $\overline{M}_{Ks}$ from \citet{liu02} (assuming the Cepheids distance modulus for the
Fornax cluster), excepting NGC~221, 224, 3379, and 4374, that are from \citet{pahr94}, and NGC~4406 and 
4472, from \citet{jens98}. 
Col.\ (9). $H\beta$ index from \citet{jens03}, excepting NGC~4278, 4374, 4486, 4649, 4552,
estimated from \citet{kobayashi99}, and NGC~4621, derived from \citet{kunt01}. 
}
\label{tab_tot}
\end{deluxetable*}

Curiously enough, note from the equation set that $K_{\rm tip}$ seems 
slightly less sensitive to $\eta$ than $\overline{K}$. As we will see in a moment, such a 
smaller variation range for $K_{\rm tip}$ stems from the ``bottoming out'' 
effect of the RGB tip, that eventually replaces the AGB in providing the brightest stars in a SSP 
when mass-loss increases to $\eta \gg 0.5$. 

The now established importance of mass-loss in old stellar populations dominated by stars with 
$M_* \simeq 1$~M$_\odot$ was first emphasized in a series of important
theoretical contributions in the early seventies \citep{cr68,ir70,cast70};  
this was urged by the intervening 
observational evidence that a mass spread was needed to reproduce HB and AGB morphology
in Galactic globular clusters \citep{dms72,saasfee,ir83}. To consistently 
match the observed AGB tip in local globular clusters, for instance, \citet{ffpr76} suggested a 
fine-tuning value of $\eta \simeq 0.4 \pm 0.1$. This calibration for old Pop~II stars might 
however not so straightforwardly be extended, {\it a priori}, to more metal-rich 
environments, as one could realistically expect mass-loss efficiency to depend on (increase with?) 
metallicity.

Depending on mass-loss strength, standard stellar evolution theory basically features 
three characteristic scenarios that constrain post-RGB evolution of low-mass stars
(see Fig.~5). Under different physical conditions and to a different
extent, each one of these cases eventually leads to the formation of hot
stars, thus potentially supplying an important contribution to the integrated 
ultraviolet luminosity of a galaxy.

{\it (a)} For $\eta \lesssim 0.4$, models tell us that at the end of RGB evolution, low-mass stars ignite 
helium in a degenerate core (the so-called ``helium flash''). A bright AGB evolution
has to be expected, reaching or even surpassing the RGB tip luminosity. If thermal pulses are set 
on ($\eta \lesssim 0.3$), stars end up loosing the external envelope and generating a planetary nebula
\citep{ir83}. In this case, hot post-AGB nuclei of nebulae easily exceed an effective temperature
of 30\,000~K \citep{schon83}.

\begin{figure}
\includegraphics[width=0.9\hsize]{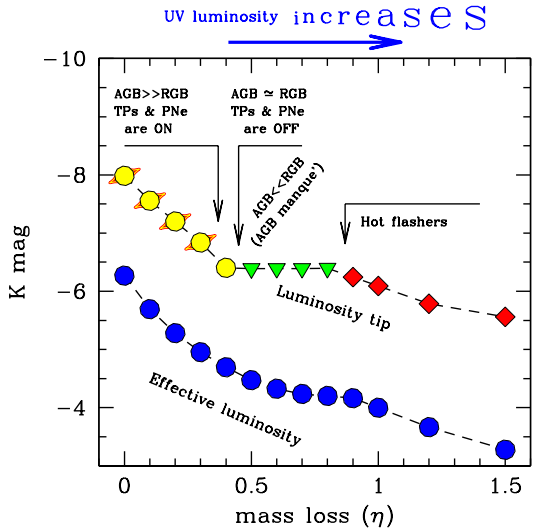}
\figcaption{Theoretical $\overline K$ and $K_{\rm tip}$ relationship with increasing mass-loss 
rate along red-giant evolution. The illustrative case of a 15 Gyr old SSP of solar 
metallicity, after \citet{buzzoni89}, is considered. Note that, for $\eta \lesssim 0.4$, the AGB luminosity tip is 
brighter than the RGB and PNe are produced; if mass loss increases, then AGB luminosity is further 
reduced (and PN formation correspondingly thwarted), until stars at the RGB tip begin to dominate 
as the brightest objects in the population ($\eta \gtrsim 0.5$). For even higher
mass-loss rates ($\eta \gtrsim 0.8$) stars undergo incomplete RGB evolution, leaving
the branch ``midway'' and igniting helium at a higher effective temperature (hot flashers). 
}
\label{sketch}
\end{figure}

{\it (b)} If $\eta \gtrsim 0.4$, the PN event is aborted \citep[][]{buzzoni06}, and AGB 
evolution is partially or fully inhibited \citep{greggio,ct,dorman}. In this case, the brightest 
stars (both in bolometric luminosity and in the infrared) belong to the RGB tip, and they settle on 
the HB with a He core mass close or
slightly exceeding 0.47~M$_\odot$ \citep[e.g.,][]{sg76,seidel,charbonnel96,piersanti04}. 
For slightly increasing mass-loss efficiency, the stellar envelope becomes thinner and thinner, 
and the hot internal core is unveiled. If this is the case, stars likely settle at high 
effective temperature ($T_{\rm eff} \gg 20\,000$~K) and originate a blue HB morphology.

{\it (c)} As a final case, for an even stronger mass-loss ($\eta \gtrsim 0.8$), the stellar
envelope might be wiped out well before the full completion of the RGB phase. 
As a consequence, standard HB and AGB evolution would be fully inhibited; stars
would be left ``midway'' along the RGB, with their rapidly exhausting H-shell 
emerging at ever shallower stellar layers. Such ``hot flasher'' stars would therefore postpone
the He ignition by moving straight to the high-temperature region of the c-m diagram, 
as is likely observed, indeed, in the cluster NGC~2808 \citep{sweigart02,castellani06}.

\begin{figure*}
\centerline{
\includegraphics[width=0.73\hsize]{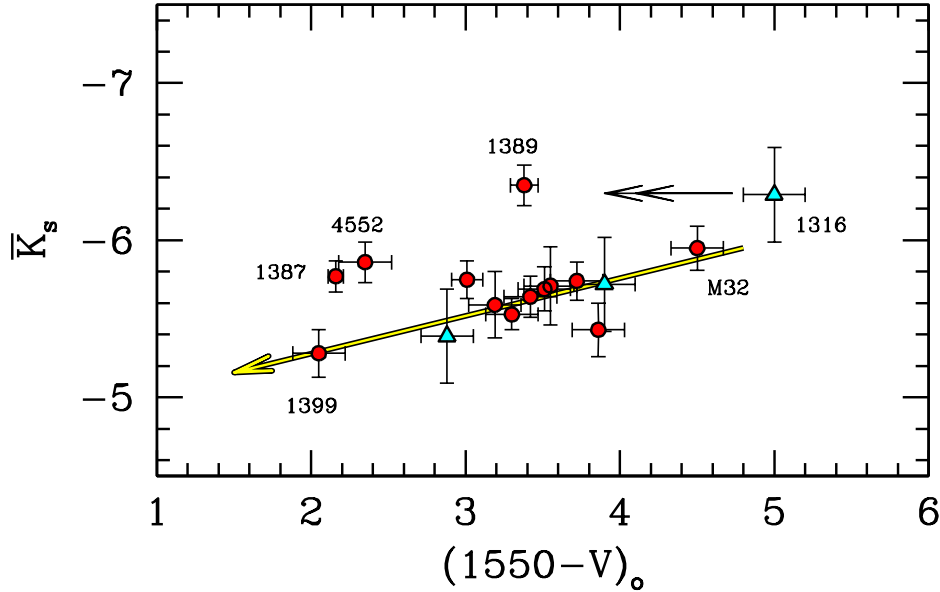}}
\figcaption{
Absolute fluctuation magnitudes vs.\ ultraviolet color (1550 - $V$)
for the galaxy sample of Table~2. Both $\overline{K_s}$ and (1550 - $V$) have been duly 
corrected for Galactic reddening. Triangles mark galaxies with $\overline{K_s}$ extrapolated 
from $\overline{F160W}$, as explained in footnote~\ref{nota}. Note the outlying cases of 
NGC~1387, NGC~4552, and NGC~1389 with an infrared effective magnitude that is $\sim 0.7$~mag 
too bright for their UV excess.
The relevant case of the merger radio-galaxy NGC~1316 (Fornax A) is also singled out in
the plot, where the arrow indicates that a ``bluer'' (1550-$V$) color might be more appropriate 
for this galaxy (with a more negligible impact on $\overline{K}_s$, though), as a consequence 
of a strong internal absorption due to the observed presence of dust lanes. Excluding these
controversial objects (discussed in more detail in Sec.~4), the data sample 
correlates fairly well ($\rho = -0.75$, see arrow), indicating a less 
deployed AGB for UV-enhanced galaxies.
}
\label{ell2}
\end{figure*}

In terms of UV contribution, cases {\it (a), (b)}, and {\it (c)} represent a sequence of 
increasing energy budget, as a prevailing fraction of stellar nuclear fuel is more
efficiently burnt at high effective temperature.
The outlined scenario has, of course, to be regarded as a general scheme, where the $\eta$
thresholds discriminating among the different 
evolutionary regimes slightly depend on SSP age (through the
MS Turn Off mass). While the illustrative case of Fig.~5 refers to a 15 Gyr SSP,
one could easily verify that the overall trend for $K$ luminosity is maintained, for instance
for a 10~Gyr population, by just ``shrinking'' the $\eta$-scale of the plot by a factor of 
$\sim 1.5\times$; this means, for example, that {\it AGB-manqu\'e} stars are produced in a 
10~Gyr scenario for $\eta \gtrsim 0.6$.

\section{UV upturn and AGB deployment in elliptical galaxies}

In order to assess the relevance of the previous theoretical framework to the appearence and strength
of the ``UV-upturn'' phenomenon in ellipticals, it is convenient to parameterize our analysis
in terms of the $(1550-V)$ color, that is a measure of the galaxy emission around the 1550~\AA\ region
vs.\ the Johnson visual band, as originally defined by \citet{burstein88}.
To this end, the \citet{burstein88} IUE galaxy sample has been taken as a reference; we supply 
in Table~\ref{tab_tot}, for each object, the infrared effective magnitudes from \citet{jens03}, \citet{liu02},
\citet{pahr94}, and \citet{jens98} and, when available, further relevant pieces of information, like
stellar velocity dispersion, integrated $(B-V)$ color, H$\beta$ and Mg$_2$ Lick indices, and the specific PN rate
per unit galaxy luminosity \cite[from][]{buzzoni06}.

As a  first relevant clue in our analysis, Fig.~6 is suggestive of a trend among most 
ellipticals, with UV-enhanced galaxies
being about one mag fainter in $K_s$ effective magnitude (some 0.7~mag in the
inferred $K_{\rm tip}$), compared to UV-poor systems.\footnote{In order to increase
the displayed galaxy sample on the different plots that involve infrared photometry, 
for some objects we derived the $K$-band effective luminosities based on the $F160W \equiv H$ 
photometry only assuming, empirically from Table 2, 
$\langle \overline{H}-\overline{K}_s \rangle = +0.9$, with a conservative error bar of 
$\pm 0.3$~mag. The extrapolated $\overline{K}$-mag data are displayed
with a different marker on the plots.\label{nota}}
On the plot one should report, however, a couple of outliers, displaying
a brighter $\overline{K}$ magnitude (or, alternatively, a ``bluer'' $(1550-V)$
color). These include NGC~1387 and NGC~1389 in the Fornax cluster, and NGC~4552 in Virgo. 
We will be back to these objects in the next section, for a brief discussion.
In addition, we have to mention that the ``merger'' radio-galaxy NGC~1316 (Fornax A)
exhibits visible dust lanes that, when accounted for, may lead to an intrinsically ``bluer'' 
(1550 - $V$) color, although with possibly negligible effects on the $K$ SBF magnitude.

After excluding these controversial objects, the data tend to display a fairly good correlation
($\rho = -0.75$) that, also recalling eq.~(\ref{eq:eta}), we can approximate with 
\begin{equation}
\Delta \overline{K}_s\ \simeq\  1.4\ \Delta K_{\rm tip}\ \simeq\ -0.2\ \Delta (1550-V).
\end{equation}

If we entirely ascribe the dimming in red-giant tip luminosity suggested by Fig.~6
to a mass-loss effect, then eq.~(\ref{eq:eta}) implies a spread $\Delta \eta \simeq 0.4$ among the
galaxy population, with UV-enhanced ellipticals requiring a Reimers parameter $\eta \simeq 0.3$-0.4,
very similar to what \citet{ffpr76} derived, empirically, for the Galactic globular clusters.
According to Fig.~5, this confirms that, overall, the AGB is deployed in luminosity 
slightly above the RGB tip, so that the brightest stars in ellipticals should indeed always be genuine
AGB stars.

\begin{figure}
\includegraphics[width=\hsize]{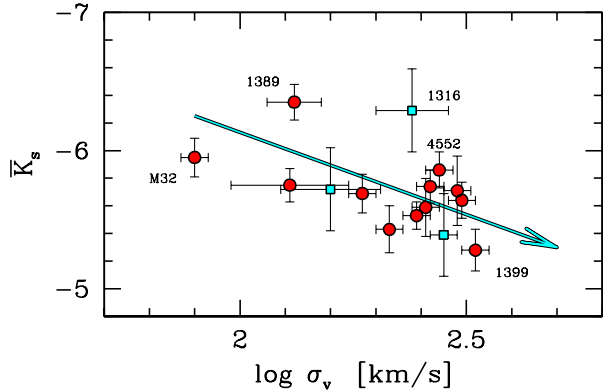}
\figcaption{
Absolute fluctuation magnitudes vs.\ velocity dispersion 
for the galaxy sample of Table~2. Squares mark galaxies with $\overline{K_s}$ extrapolated 
from $\overline{F160W}$, as explained in footnote~\ref{nota}.
}
\label{fig201}
\end{figure}

\begin{figure}
\includegraphics[width=\hsize]{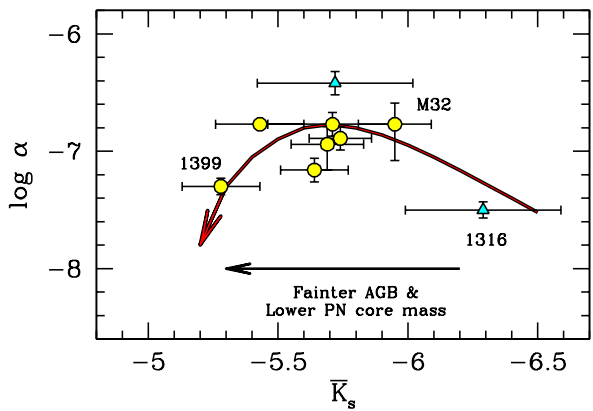}
\figcaption{The luminosity-specific PN number ($\alpha$) vs.\ effective $K$ 
luminosity for the galaxy sample of Table~2. The triangles mark galaxies with $\overline{K}_s$ 
inferred from $\overline{F160W}$ (see footnote~\ref{nota}). Note that a fainter AGB roughly correlates with
a scantier PN stellar population among old/metal-rich (UV-enhanced) ellipticals, as expected
from the long post-AGB to PN nucleus transition time, combined with the (shorter) evaporation timescale of the ejecta
\citep[compare the sketch on the plot with Fig.~15 in][, for a full discussion]{buzzoni06}. 
A correspondingly low PN rate per unit galaxy luminosity is also to be expected, on the other hand,
for younger stellar populations, as in the case of NGC~1316, as a consequence of higher stellar masses
and a reduced PN nuclear lifetime.
}
\label{fig200}
\end{figure}

There is an obvious caveat, though, in this simplified picture.
If a standard monolithic scenario is assumed for early-type galaxy formation, then gravitational 
collapse of the primeval gas clouds should have proceeded over a free-fall timescale  
$\tau_{\rm ff} \propto \sigma_v^{-1} \propto {\rm M}_{\rm gal}^{-1/2}$ \citep[e.g.,][]{larson74},
leading  to a mass--age--metallicity relation for early-type galaxies, 
where more massive galaxies are both older and more metal-rich
than less massive ones \citep[e.g.,][]{pahre98}.
The combined action of age and metallicity plays an important role to modulate
the red-giant luminosity tip, as the AGB is ``naturally'' brighter among young 
or metal-rich SSPs (see, again, Fig.~3).

The complex interplay of the different galaxy physical properties is well depicted in 
Figs.~7 and 8. For one, as \citet{burstein88} had already shown, the 
UV upturn seems to appear among the most massive and metal rich ellipticals (i.e.,
there is a positive correlation with  $\sigma_v$ and the Mg$_2$ index, respectively). 
Figure~7 indicates for these galaxies also a less extended AGB, as for a lower stellar 
core-mass distribution.
This feature is likewise expected to directly affect the PN rate per unit galaxy 
luminosity (the so-called ``$\alpha$'' index, see Fig.~8); { indeed, there is
a scantier} PN stellar population 
among old/metal-rich (UV-enhanced) ellipticals \citep{ferguson93,buzzoni06}.

\begin{figure*}
\centerline{
\includegraphics[width=0.73\hsize]{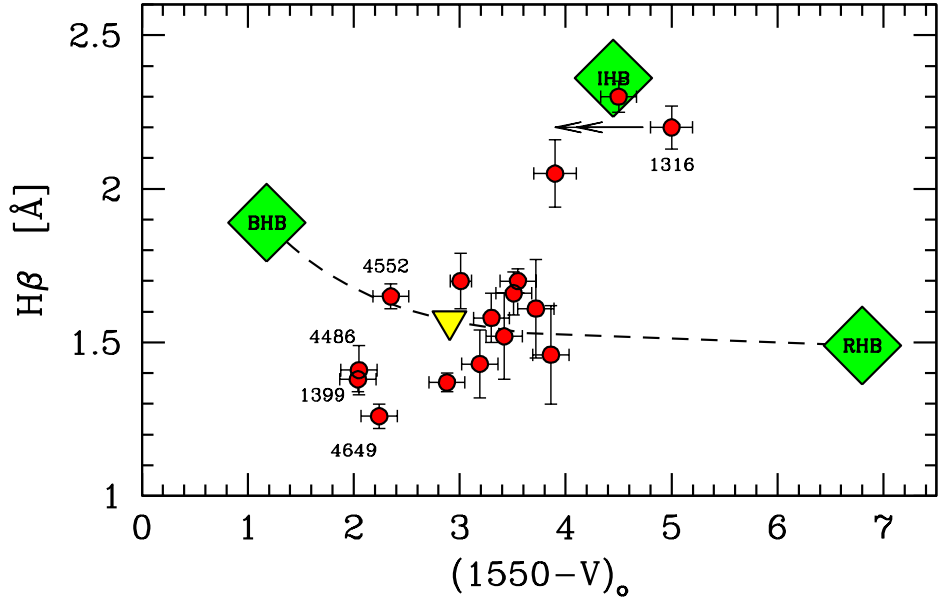}
}
\figcaption{
Lick H$\beta$ index vs.\ (reddening corrected) UV color $(1550 - V)_o$, 
for the galaxy sample in Table~2. Three reference SSP models are superposed, 
after \citet{buzzoni89}, exploring different HB morphologies (namely a red, RHB, 
an intermediate, IHB, and a blue, BHB, temperature distributions, 
as explained in the text), assuming a 15 Gyr, 
slightly metal-rich ([Fe/H] = +0.2) and with a fixed Reimers mass-loss parameter 
($\eta = 0.3$) stellar population. Note that the bulk of ``UV-upturn'' ellipticals
need a mixture of blue and red HB stars, roughly in a proportion of [N(RHB):N(BHB)] = [80:20],
as marked by the big triangle in the plot. For this composite stellar population, Fig.~10 reports the
resulting synthetic c-m diagram and the integrated SED.
}
\label{hbeta}
\end{figure*}

\subsection{Constraints on HB morphology}

Observations of local star samples  show that the H$\beta$ index reaches its 
maximum strength for A-type stars in the temperature range 
T$_{\rm eff} \simeq 8\,000 \to 10\,000$~K \citep{buzzoni94}.
Given this selected sensitivity, synthesis models of old SSPs \citep[][]{buzzoni94,maraston01}
predict the integrated index to be strongly enhanced (by roughly 0.8~\AA\ or more) in the presence 
of a broad color-extended HB, as observed for most metal-poor Galactic globular clusters.
This important piece of information could therefore usefully complement the more extreme $(1550-V)$ 
color in the analysis of HB morphology for unresolved stellar populations in distant galaxies.

The situation is summarized in Fig.~9, where elliptical galaxy data are compared with
three illustrative cases of 15 Gyr SSPs with slightly supersolar 
metallicity ([Fe/H]=+0.3), and different ranges of HB temperature distribution.  
In particular, in the figure we account for 
{\it (i)} a red HB (RHB) morphology, that is, a clump of red stars very close to the RGB location, 
mimicking the real case of metal-rich Galactic globular clusters, like 47 Tuc;
{\it (ii)} an intermediate HB (IHB) morphology, with a broad (roughly bell-shaped) temperature 
distribution peaked about the A-stars' temperature range and extending up to $T_{\rm eff} \sim 12\,000$~K; 
{\it (iii)} a blue HB (BHB) morphology, peaked at $20\,000$~K and with a tail of hot sdB stars, 
up to a temperature of $40\,000$~K \citep[see][, for further details]{buzzoni89}.

Interestingly enough, the comparison with the observations shows that ``UV-upturn'' ellipticals
{\it cannot} be compatible with any BHB morphology alone. Too many hot stars would make the 
1550~\AA\ galaxy luminosity exceedingly bright compared to the energy bulk released at the $V$ band. 
On the other hand, neither an IHB is viable, given the catastrophic impact of its A-star 
component on the integrated $H\beta$ index of the population \citep[e.g.][]{rose99}. 
Definitely, an optimum match to these systems requires a mixed contribution, where the bulk 
of the galaxy luminosity (over 80\%) is provided by ``standard'' (metal-rich) stellar populations 
with a red HB, upon which a 20\%, or less, in terms of total optical contribution,
BHB component is superposed.
Figure 10 provides an illustrative picture of this case, showing the synthetic c-m
diagram and the integrated SED for the composite stellar population corresponding to the
big triangle in Fig.~9.

\begin{figure*}
\centerline{
\includegraphics[width=0.41\hsize]{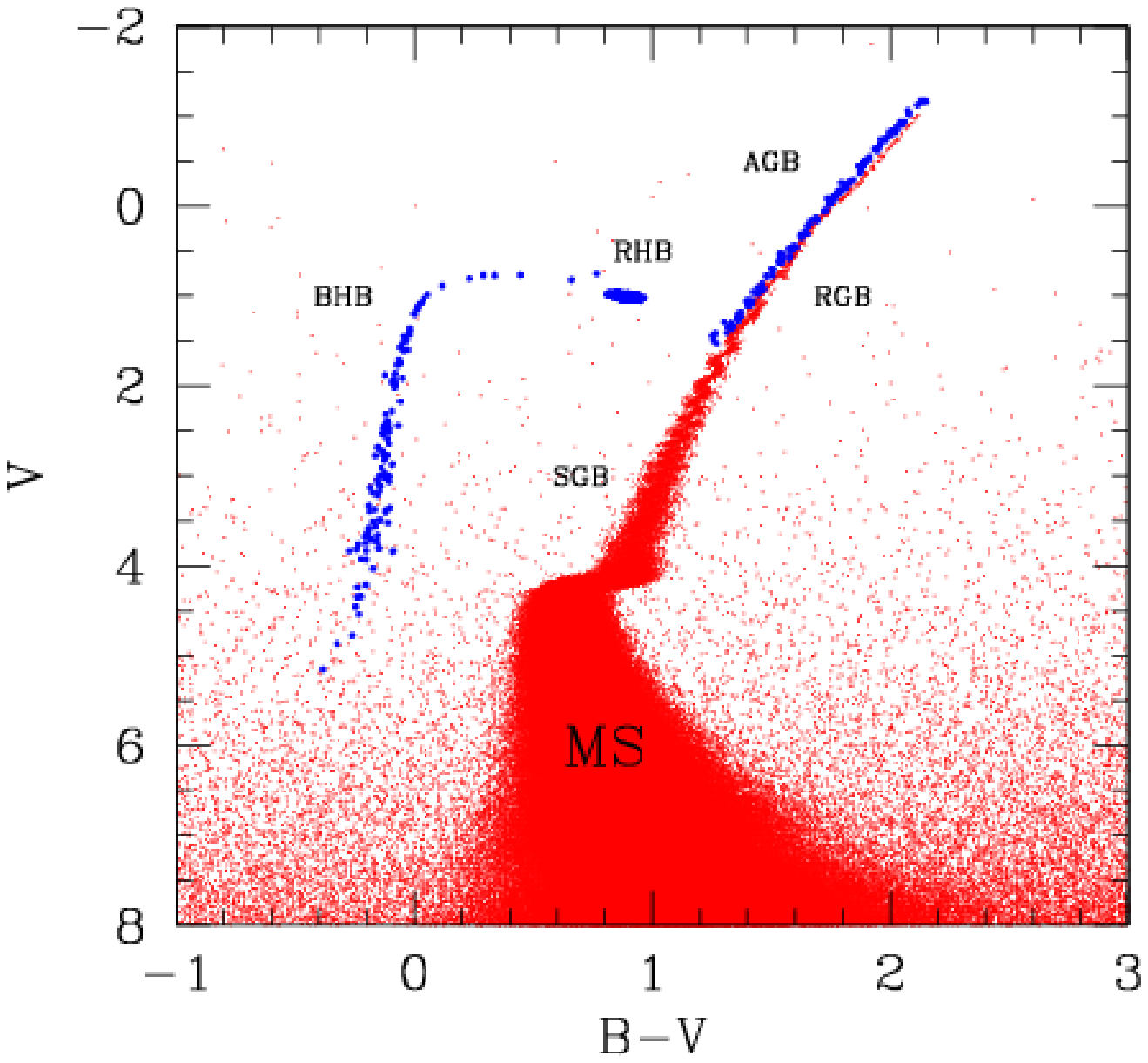}
\includegraphics[width=0.59\hsize]{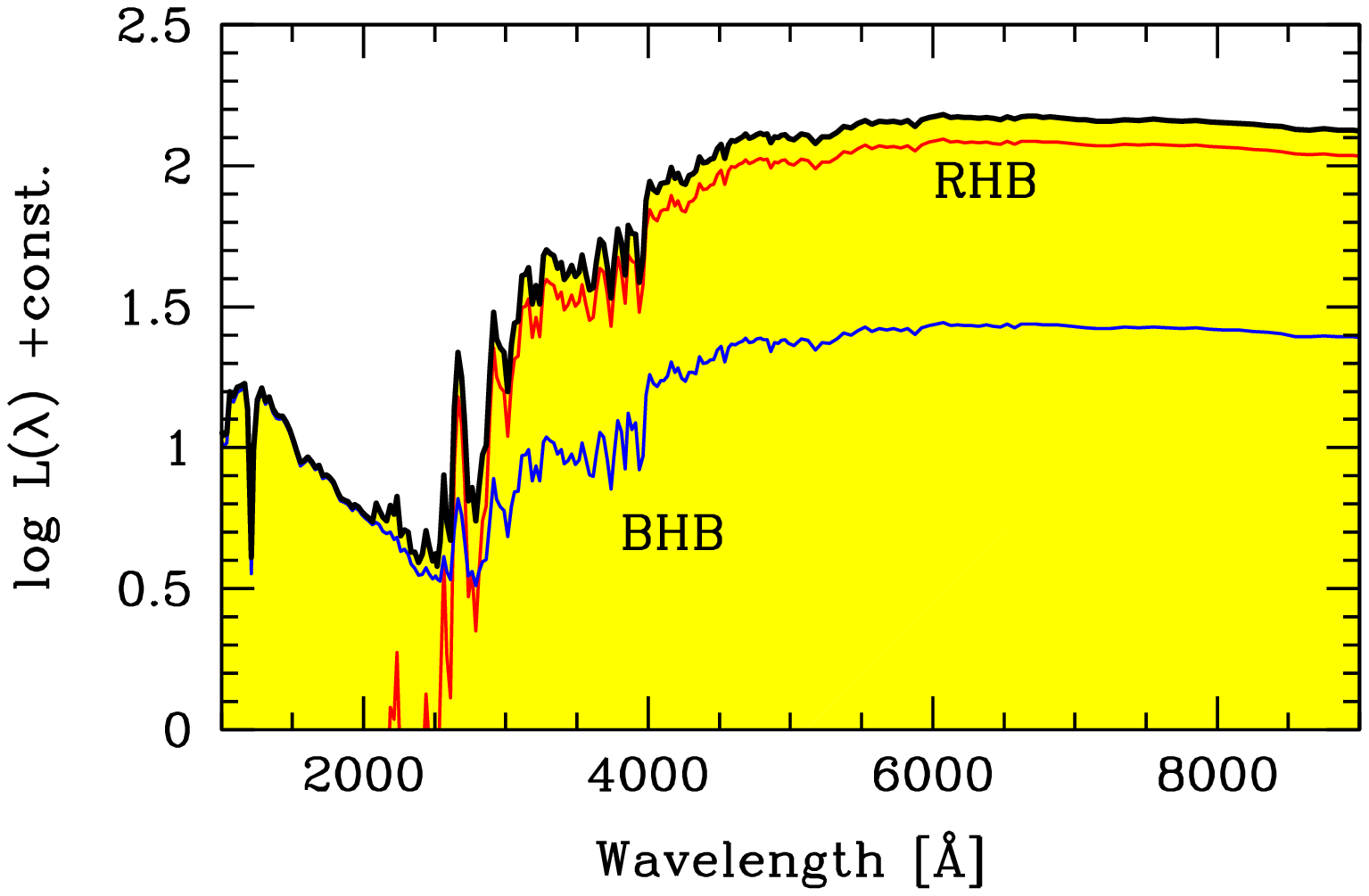}
}
\figcaption{{\it Left panel -} Synthetic c-m diagram for the reference composite stellar
population matching ``UV-upturn'' ellipticals, as discussed in Fig.~9 (the corresponding big
triangle in the plot), after \citet{buzzoni89}.
The resulting mix assumes a bimodal HB morphology, with a prevailing bulk of red HB (RHB) stars and 
a hot tail (BHB), extending up to $T_{\rm eff} \simeq 40\,000$~K. The RHB stellar component provides 
about 80\% of the total luminosity. An age of 15 Gyr is assumed in all cases, with a moderately metal-rich chemical composition 
(i.e.\ [Fe/H] = +0.2), a Salpeter IMF, and a fixed Reimers mass-loss parameter $\eta = 0.3$.
A Poissonian error is artificially added to the data to better appreciate the number density distribution
of stars along the different evolutionary branches of the diagram.
The integrated SED of the whole population is displayed in the {\it right panel}, disaggregating
the luminosity contribution from the two star samples.
}
\label{cmdhb}
\end{figure*}

So, in full agreement with \citet*{bn04} conclusions, a somewhat dichotomic scenario seems to emerge 
for giant ellipticals, where the occurrence of the ``UV-upturn'' phenomenon is intimately 
related to the appearence, among the galaxy HB population, of a hot and very low-mass 
stellar component \citep[M$_* \ll 0.58$~M$_\odot$, e.g.][]{sg76}, that at some stage 
begins to join the standard red-clump stars.
It could be useful to note, in this regard, that two metallicity ranges seem to 
favor the formation of EHB stars, through an optimum combination of MSTO stellar 
masses and subsequent mass-loss via stellar winds along the RGB. As displayed in Fig.~11,
[Fe/H] values of either $\simeq -0.7$~dex or $\gtrsim +0.5$ provide in principle the
needed low mass at the onset of HB evolution in a standard Reimers mass-loss framework
with $\eta \gtrsim 0.5$.

To add even further to this challenging situation, however, one also has to report that 
high-resolution UV spectroscopy of a sample of six UV-strong ellipticals (namely, NGC 1399, 
3115, 3379, 4472, 4552, and 4649) indicates for the claimed BHB component a moderately 
sub-solar ($Z \sim 0.1~Z_\odot$) metallicity \citep{brown97}. If confirmed,\footnote{Surface metal 
abundance can easily be biased by diffusion mechanisms 
in the atmosphere of sdB stars \citep{unglaub01}.}
this result might lead to identify the {\it metal-poor} stellar component of giant ellipticals
as the one responsible for the UV excess.
Whatever the exact mechanism, this will cause HB morphology to readily assume a bimodal  
distribution, with a relative lack of warm stars at the intermedate $T_{\rm eff}$ range
pertinent to the A spectral type (see Fig.~10). 

\clearpage

\begin{figure}
\centerline{
\includegraphics[width=\hsize]{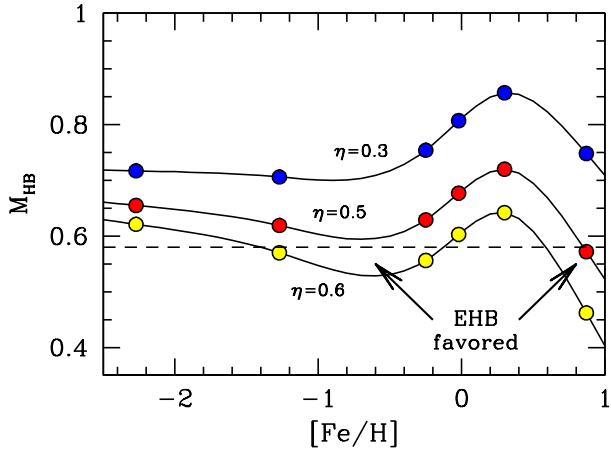}
}
\figcaption{The expected representative stellar mass at the onset of the HB evolution
for Buzzoni's (1989) 15 Gyr SSP models with different metallicity and mass-loss rate, according to
a Reimers parameterization ($\eta$). In combination with a moderately enhanced 
($\eta \simeq 0.5$ or higher) mass loss, two preferred ranges of metallicity (namely
$[Fe/H] \simeq -0.7$~dex and $\gtrsim +0.5$) may more easily favor the presence of low HB 
masses ($M_{HB} \lesssim 0.58$~M$_\odot$, see dashed line in the plot) and the corresponding 
appearence of a hot-temperature tail in the HB morphology.
}
\end{figure}

\section{Summary and conclusions}

In this paper we have carried out a synoptic analysis of the different observational features 
that have to do with the UV luminosity excess in elliptical galaxies.
As far as the canonical picture is assumed, with old stellar populations dominating early-type 
galaxy luminosity, the appearence of the ``UV upturn''  should readily call for a profound 
change in the c-m diagram of galaxy stellar populations, not only involving the hot 
stellar component of the galaxy but also reverberating on red-giant evolution at
the low-temperature regime.
As we mainly deal with distant, unresolved stellar populations, our analysis has to rely on
a combined approach, matching infrared and ultraviolet diagnostic tools in order to probe 
the main features of the stellar c-m diagram, starting with integrated galaxy photometry.

Theory of surface-brightness fluctuations provides, in this sense, a natural and quite
powerful way to go deep inside the problem and, as far as the infrared wavelength interval
is considered, we have demonstrated theoretically that a straight and very clean relationship 
is in place between a macroscopic measure, such as the galaxy fluctuation magnitude, and 
the corresponding individual magnitude of the brightest stars in turn at the tip of the 
red-giant (AGB+RGB) phases (Fig.~2).

Played in the $K$ band, this correlation leads, from a measurement of $\overline{K}$, to a 
value for $K_{\rm tip}$: 
\begin{equation}
K_{\rm tip}  = 0.75\ \overline{K} - 3.1,
\label{eq:calib}
\end{equation}
with a $\pm 0.2$~mag internal uncertainty. 
As we showed in Sec.~2, our SSP theoretical predictions find full support from the observations,
and a direct check on the MC star clusters confirms the $\overline{K}$ vs.\ $K_{\rm tip}$ 
relationship to be a much more 
general and deeply intrinsic property of stellar populations, virtually independent from 
any assumption about age, metallicity, IMF, and mass loss parameters.
Given its nature, this relationship cannot, by itself, help disentangle the problem of 
age/metallicity degeneracy; however, quite fruitfully, it provides us with a very direct probe 
of AGB properties, in a number of relevant details that directly deal with the mass-loss 
impact and the mass of dying stars (Fig.~4).

Our effort toward exploring the infrared side of galaxy SEDs has a twofold aim since, as a consequence
of the basic principle of energy conservation, any gram of stellar fuel spent to feed 
ultraviolet luminosity cannot (and will not) be spent at longer wavelengths.  
This has led to the key issue of this paper, summarized in
Fig.~6, that the {\it strengthening of the UV rising branch is always seen to correspond 
to a weakening in the AGB luminosity extension}, as traced by galaxy $K$ fluctuation magnitude.

This ``shortening'' in AGB deployment is mainly recognized among giant ellipticals
($\overline{K}$ becomes fainter with increasing galaxy velocity dispersion, $\sigma_v$,
see Fig.~7), and could mainly be ascribed to an age effect, as the AGB tip naturally fades 
in luminosity with increasing age of the system (Fig.~3), and high-mass galaxies are recognized 
to be older than systems of lower mass \citep[e.g.][]{burstein88,bressan96,liu02,jens03,gonz05a,renziniaraa}.
However, the relationship in place likely calls for a more elaborated physical scenario, once 
the full range of observing evidences is added to our analysis.

\begin{figure}
\centerline{
\includegraphics[width=\hsize]{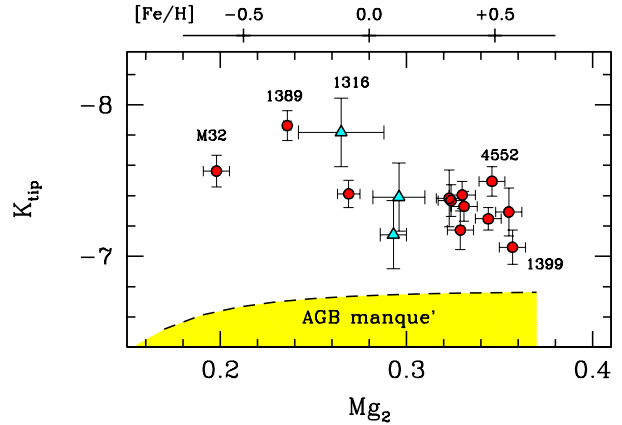}
}
\caption{
$K_{\rm tip}$ vs.\ Lick Mg$_2$ index, for the galaxy sample in 
Table~\ref{tab_tot}. The Lick index is assumed to trace galaxy metallicity according to
the \citet{buzzoni92} calibration, as reported on the top axis of the plot.
The observed decrease in $K_{\rm tip}$ with increasing $[Fe/H]$ can be mostly explained if 
more metal-rich galaxies are also older, as in a standard monolythic scenario for galaxy formation. 
The dashed line on the plot marks the minimum luminosity required for stars to
experience the thermal pulsing phase along their AGB evolution, and thus end their evolution
as PNe.
}
\label{eta}
\end{figure}

{\it (a)} Besides being old, ``UV upturn'' galaxies are also metal rich (i.e., stronger in 
Mg$_2$ Lick index). Disregarding any change in mass-loss rate, stellar tracks predict slightly 
more massive stars to evolve off the MS at a fixed age, with increasing metallicity 
\citep[e.g.,][]{bressan94,cantiello03}. This leads to correspondingly more massive AGB stars 
and a brighter AGB luminosity. Facing the observed trend in galaxy distribution, as summarized 
in Fig.~12, metallicity effects evidently enter by mitigating the dimming action of age on 
$K_{\rm tip}$ with increasing galaxy mass. In any case, the interplay between age and metal 
abundance actually makes the derived range for the Reimers mass-loss parameter 
(i.e.\ $\Delta \eta \simeq 0.4$, as discussed in Sec.~3) a safe upper limit. In fact, it suggests 
that metal abundance does {\it not} modulate by orders of magnitude
mass-loss efficiency via stellar winds.\footnote{Observational evidence about the link between 
metallicity and mass-loss in the Milky Way and the Magellanic Clouds is
contradictory. For example, \citet{groe95} find indications that mass-loss rate
(not necessarily mass-loss {\em efficiency}) in single AGB stars is linearly proportional to $Z$. 
Conversely, also from data of single stars, \citet{vanl00} argues that $\dot M$ is 
metallicity-independent. 
From a theoretical point of view, \citet{cantiello03} models suggest that, if
mass-loss is really proportional to metallicity, its effect to dim near-IR
effective luminosities on average almost exactly offsets the brightening effect 
of metallicity itself.}

\begin{figure}
\centerline{
\includegraphics[width=\hsize]{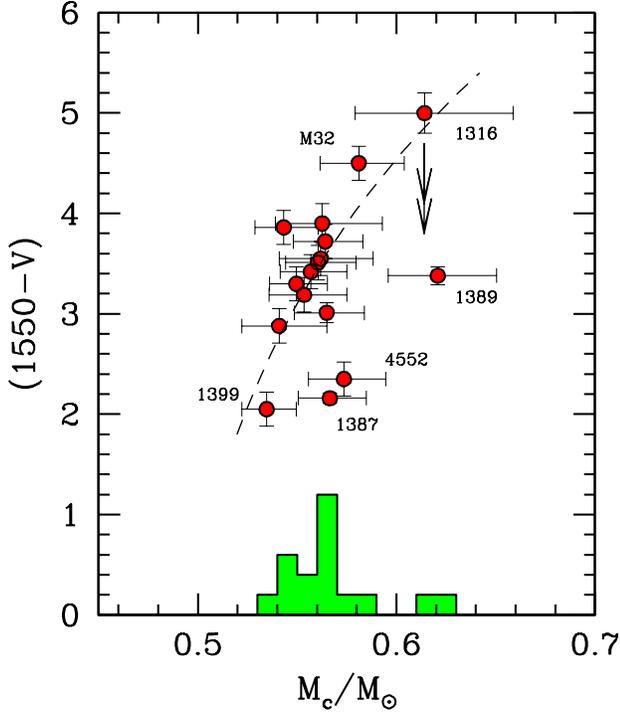}
}
\figcaption{Observed ultraviolet color $(1550-V)$ vs. core mass of stars 
at the AGB luminosity tip, as inferred from eq.~(\ref{mctip}), for the
elliptical galaxy sample of Table~2. The M$_c$ distribution is summarized
in the lower histogram, and is the maximum actual mass allowed to luminous 
stars in the galaxies. One sees that mass of dying stars tends to decrease 
with increasing UV-upturn strength, being in general $\lesssim 0.57$~M$_\odot$
among giant ellipticals. The ``outlier'' objects of Fig.~6 are 
identified again here, with galaxies labelled according to their NGC number.
The arrow for NGC~1316 accounts for the claimed strong internal reddening for
this galaxy.
Displayed uncertainties for the derived values of M$_c$ take in the full error 
budget-- each component of $\sigma(M_c)$ being added in quadrature--, including 
the contribution of $\overline{K}$ observations, $K$-band bolometric correction 
($\sigma = \pm 0.2$~mag), and the $K_{\rm tip}$ vs. $\overline{K}$ calibration 
($\sigma = \pm 0.2$~mag). See text for a discussion.
}
\label{mcore}
\end{figure}

{\it (b)} A match of galaxy $\overline{K}$ data with the calibration of Fig.~5 confirms 
that, all the way, brightest stars in ellipticals are genuine AGB members, reaching the thermal-pulsing phase 
(see also Fig.~12), and with the AGB tip exceeding the RGB tip by some 
0.5-1.5~mag. In the temperature range of M giant stars, a major fraction of 
bolometric luminosity is emitted through the $K$ band, and bolometric correction is a nearly 
constant quantity that we can estimate from $({\rm Bol} - K) = +2.75 \pm 0.2$~mag \citep{johnson}; 
we can therefore straightforwardly translate the galaxy fluctuation magnitude into an estimate 
of the bolometric tip luminosity, $L_*^{\rm tip}$, and therefrom of the corresponding 
stellar core mass.\footnote{A somewhat linear relationship
between stellar luminosity and core mass is a general consequence of any evolutionary stage characterized by a (multi) shell-burning
regime in the presence of a relatively thin external envelope \citep{paczynski}. 
This is actually the case of both pre-He flash evolution along the RGB and the thermal pulsing 
phase along the AGB \citep[see][, for a more general discussion]{ir83,boothroyd}.}
From our previous calibration (eq.~\ref{eq:calib}), we can write then 
\begin{equation}
\log L_*^{\rm tip} = -0.4[(0.75\overline{K}-3.1) +2.75 -4.72]
\end{equation} 
(where the Sun has magnitude M$_{{\rm Bol},\odot} = +4.72$).
Following \citet{boothroyd}, from the assumed core mass-luminosity relation for the solar 
metallicity range, this leads to
\begin{equation}
M_c = {{L_*^{\rm tip}}\over 52\,000}+0.456 \qquad\quad [M_\odot]. 
\label{mctip}
\end{equation}
Figure~13 reports the inferred core-mass distribution, at the PN onset,
for our galaxy sample.

Note that this is the {\it maximum} actual mass allowed to luminous stars in each
galaxy environment, and demonstrates that the {\it mass of dying stars tends to decrease 
with increasing UV-upturn strength}, being in general $M_{\rm dying} \lesssim 0.57$~M$_\odot$
among giant ellipticals. For this mass range, PN lifetime is the largest possible, 
but the timescale for the nebula to be visible is critically constrained by the 
transition time ($\tau_{\rm tt}$) needed by the post-AGB stellar core to be hot enough to 
``fire up'' the ejected envelope and become a hard UV emitter \citep[][]{letizia00,marigo05}. 
The evident drop of $\alpha$ among strong UV-upturn galaxies (Fig.~8) might be a direct
consequence, therefore, of an increasing blocking effect of $\tau_{\rm tt}$ along the 
inferred $M_c$ range \citep[i.e., the stellar core takes longer to heat-up than the shell 
to evaporate;][]{buzzoni06}, to which one has to further add a size cut in 
the overall PN population, as a result of the EHB progenitors (M$_* \lesssim 0.52$~M$_\odot$) 
evolving as {\it AGB-manqu\'e} stars and therefore skipping the nebula event.

{\it (c)} We remarked, in Sec.~3, the importance of the integrated $H\beta$ index as a fairly 
selective tracer of the warm (T$_{\rm eff} \simeq 8\,000-10\,000$~K) stellar component 
in the galaxy stellar population. A proper assessment of the photometric contribution from 
this range of temperatures is of paramount importance in the framework of early-type galaxy evolution,
in order to single out any signature of recent (i.e., in the last few Gyr or so)
star formation or, conversely, of intervening evolution of the HB morphology among old 
SSPs, as in a more standard canonical scenario.

As far as UV-upturn galaxies are concerned, the study of $H\beta$ distribution clearly  
points to {\it a substantial lack of A-type stars} in the galaxy mix (see Fig.~9).
While, on one hand, this definitely secures the ``quiescent'' nature of these galaxies, 
it also poses, on the other hand, a stringent constraint on HB morphology in their old-age 
context. In fact, a bimodal temperature distribution
is required for the HB to assure, at a time, both an enhanced UV emission {\it and} a
conveniently low $H\beta$ feature.
Thus, in these systems the expected prevailing bulk of red  HB stars should be 
accompanied, at some point, by a residual population of blue (metal-poor?) HB objects 
(coincident with EHB stars in the current empirical classification scheme), peaked 
at about 20\,000-40\,000~K, in a proportion of, roughly, $N_{\rm RHB}:N_{\rm BHB}~\simeq [80:20]$. 

On the other hand, to complete the picture, one cannot neglect the masking effects of age distribution, 
facing a recognized evidence for low-mass ellipticals to display a more silent but 
also more continuous star 
formation along their entire galaxy life, that naturally feeds the A-star contribution, thanks       
to the bluer MS turn-off point (MSTO) exhibited by SSPs in
the $~1-3$~Gyr age range, and leads to a younger ``average" age, compared to high-mass
systems.
This is what we observe, for instance, among the resolved stellar 
populations of the Local Group dwarf spheroidals \citep{mateo98}
\citep[see, in this regard, the illustrative location of M32 in Fig.~9, and also consider the
discussion by][]{schiavon04}.

{\it (d)} Along with our discussion of the $\overline{K}$ vs.\ $(1550-V)$ relation, we
noticed, in Fig.~6, the presence of a few outliers about 0.7~mag brighter at infrared
magnitudes, or alternatively $\sim 1.5$~mag ``bluer'' in the $(1550-V)$ color, than the main galaxy 
population. In order to further investigate this issue, we tracked the relevant objects 
also in other figures, whenever possible. 

If galaxy mass ({\it alias} $\log \sigma_v$) is considered as the leading physical parameter to 
compare outlier location with respect to the bulk of the galaxy distribution (see, for instance Fig.~7), 
one must conclude that both NGC~4552 and NGC~1389 seem to have a brighter 
AGB tip rather than a bluer $(1550-V)$ color. With regard to NGC~1389, \citet{liu02} 
find that the $\overline{K_s}$ SBFs of NGC~1389 are also too bright compared to its 
($V - I$) color, a fact that would be consistent with either a higher than 
average metallicity given the age of
its most recent burst of star formation or a longer lifetime of its TP-AGB stars 
\citep{mouhcine05}. In the case of NGC~4552, however, \citet{jens96} contribute
an interesting piece of information. These authors measure near-IR SBFs 
for several galaxies in Virgo, and their results are systematically fainter 
than those obtained by \citet{pahr94}. Unfortunately, the two groups
use slightly different filters (Jensen et al.\ employ $K^\prime$, vs.\ $K_s$~of 
Pahre \& Mould), but the discrepancy is larger than can be ascribed to the 
effect of the filters.\footnote{\citet{jens96} attribute the 
difference to the higher S/N ratio of their data; the S/N ratio of the \citet{liu02} images
is similar to that of Jensen et al.'s.}  
In particular, Jensen et al.\ find 
$\overline{K^\prime} = -5.51$ for NGC~4552; assuming $\overline{K^\prime} \equiv \overline{K_s}$,       
NGC~4552 would no longer be deviant in the $\overline{K_s}$ vs.\ ($1500 - V$) plane.
Concerning NGC~1387, the dearth of
data for this galaxy in the literature (see Table~2) makes it hard to
propose an origin for its departure from the  
$\overline{K_s}$ vs.\ ($1500 - V$) sequence.\footnote{Note that NGC~1387 complies with the
$\overline{K}_s$ vs.\ ($V - I$) correlation determined by
\citet{liu02}.} On the other hand, NGC~1387 and NGC~1389 are lenticular
galaxies, like the merger remnant NGC~1316, but so are NGC~3384 and NGC~4406,
both of which do not deviate from the correlation.\footnote{The other galaxies
in our sample are all genuine ellipticals \citep{rc3}, except for NGC~224 ({\it alias} M~31),
which is so close, however, that bulge SBFs can be measured without any 
significant contamination from the disk.}
At any rate, NGC~1387 and NGC~1389 constitute privileged 
candidates for any future ``in-depth'' investigation. 

\acknowledgments
We would like to thank Gustavo Bruzual for providing us with his latest SSP models, 
in advance of publication, and Livia Origlia, for useful discussions. 
The anonymous referee is also acknowledged for his/her competent suggestions, that 
greatly helped refine the main focus of the paper.
Partial financial support is acknowledged from the Italian MIUR, under grant 
INAF-PRIN05 1.06.08.03, and Mexican CONACyT, 
under grant no.\ 48589-F, and DGAPA, under grant IN111007.

\end{document}